\documentclass[preprint,12pt]{elsarticle}

\usepackage{lineno,hyperref}
\modulolinenumbers[5]

\usepackage{amssymb}
\usepackage{amssymb}
\usepackage{graphicx}
\usepackage{amsmath,amsfonts,amssymb}
\usepackage{epsfig,makecell,float}
\usepackage{setspace,mathrsfs}
\usepackage{tocloft}
\usepackage{color}
\usepackage{textcomp}
\usepackage{multirow,indentfirst,times,color}
\usepackage[caption=false,font=footnotesize]{subfig}
\usepackage{booktabs}
\usepackage{threeparttable}
\usepackage{amsthm}
\usepackage{extarrows}
\newtheorem{thm}{Theorem}
\newtheorem{defn}{Definition}
\newtheorem{remark}{Remark}
\newtheorem{lem}{Lemma}

\usepackage[titletoc]{appendix}
\usepackage{epstopdf}
\usepackage{natbib}
\biboptions{numbers,sort&compress}

\begin{document}

\begin{frontmatter}

\title{Unlimited Sampling Theorem Based on Fractional Fourier Transform}

\author{Hui Zhao$^{a,b}$}
\author{Bing-Zhao Li$^{a,b}$\corref{mycorrespondingauthor}}
\cortext[mycorrespondingauthor]{Corresponding author}\ead{li\_bingzhao@bit.edu.cn}

\address{$^{a}$School of Mathematics and Statistics, Beijing Institute of Technology, Beijing 100081, China}
\address{$^{b}$Beijing Key Laboratory on MCAACI, Beijing Institute of Technology, Beijing 100081, China}


\begin{abstract}
	The recovery of bandlimited signals with high dynamic range is a hot issue in sampling research. The unlimited sampling theory expands the recordable range of traditional analog-to-digital
	converters (ADCs) arbitrarily, and the signal is folded back into a low dynamic range measurement, avoiding the saturation problem. We study the unlimited sampling problem of high dynamic non-bandlimited signals in the Fourier domain (FD) based on the fractional Fourier transform (FRFT). First, a mathematical signal model for unlimited sampling is proposed. Then, based on this mathematical model, the annihilation filtering method is used to estimate the arbitrary folding time. Finally, a novel unlimited sampling theorem in the FRFD is obtained. The results show that the non-bandlimited signal can be reconstructed in the FD based on the FRFT, and it is not affected by the ADC threshold.	
\end{abstract}


\end{frontmatter}


\section{Introduction}
\label{Intro}
In signal processing, sampling \cite{Communication1949,Certain2002, Sampling1992} is the primary task faced in the process of digitizing the signal.  Since Shannon's sampling theorem \cite{Unser2000} was proposed, sampling theory has been developed for more than 70 years, and its theoretical results \cite{Quantitative1999,Seven2014,Higher2019,New2007} are so rich that it has become one of the research hotspots in the field of signal processing. From a practical standpoint, point-wise samples of the function is obtained using  the analog-to-digital converter (or
the ADC), but the ADC \cite{Amplitude2009} has a limited dynamic range $[-\lambda,\lambda]$. Whenever the signal exceeds a certain preset threshold $\lambda$, the ADC will saturate, and the aliased signal will be clipped due to clipping \cite{Iterative2014,Restoring1991}. Since most signals in practical applications are not limited by broadband, the dynamic range is very wide, so self-reset ADC (S-ADC) was proposed \cite{W. Kester2009,D. Park2007,Y. Joo2010}. Each time the input signal reaches the upper (lower) saturation limit, these S-ADCs will be reset to the corresponding other thresholds, which allows the S-ADC to reset rather than saturate, resulting in analog sampling. When the signal reaches the upper (lower) threshold point, it will fold backward (front) by an integer multiple of $2\lambda$. This phenomenon is equivalent to modulo arithmetic on the input signal, which is very helpful for processing high dynamic range signals. \


Because of the S-ADC's ability to process high dynamic range signals, Bandari et al. \cite{On2017} recently made the first pioneering contribution. He proposed the unlimited sampling theorem and developed the first provable refactor of the guaranteed algorithm. The results of \cite{On2017} have led to a lot of follow-up work, and the theoretical research of unlimited sampling has gradually enriched \cite{ Sparse2019,A.Bhandari2018,Shah2018, Recovering2019, A.Bhandari2020}.
 In this modular sampling framework, many scholars have studied the sampling and reconstruction of bandlimited functions and smooth functions under different backgrounds.  The paper \cite{Above2019} is shown that the bandlimited function is uniquely characterized by modular samples under certain conditions. Ordentlich et al. \cite{Ordentlich2018} studied the recovery of quantization modulus samples by using edge information. In \cite{Generalized2018}, it gave the modulus sampling theory of S-ADC sparse signal. The unlimited sampling method based on wavelet is suitable for general smooth signals, not limited to bandlimited signals in \cite{S. Rudresh2018}. The paper \cite{A.Bhandari2021} is mainly applicable to bandlimited signals in the Fourier domain (FD) on  unlimited sampling method.\

Most of unlimited sampling frameworks are based on the bandlimited signals, but there are few articles on non-bandlimited signals. For various applications of the non-bandlimited signal models, the original results are not directly applicable. According to \cite{X. G. Xia1996}, as a generalized form of the Fourier transform (FT), the fractional Fourier transform (FRFT) can expand the signal range applicable to traditional sampling theory, because the non-bandlimited signal in FD may be bandlimited in the fractional Fourier domain (FRFD). Therefore, the application of traditional sampling theory to non-bandlimited signals cannot achieve optimal results, so it is very necessary to study the unlimited sampling theory of non-bandlimited signals.\ 

Based on the above reasons, we put forward the problem of unlimited sampling of high dynamic non-bandlimited signals in the FD based on the FRFT. We innovate the content of \cite{A.Bhandari2021}, and propose a new FRFD sampling theorem of high dynamic non-bandlimited signals in the FD based on an unlimited sampling framework. Firstly, a mathematical signal model for unlimited sampling is proposed. Secondly, based on this mathematical model, the annihilation filtering method is used to estimate the arbitrary folding time. Finally, a novel unlimited sampling theorem in the FRFD is obtained. 

This paper is organized as follows. In Section \ref{Preli}, we describe briefly the FRFT, unlimited sampling theorem in the FD. In Section \ref{Mathematical}, a mathematical signal model for unlimited sampling is proposed. The FRFD sampling theorem on unlimited sampling is proposed in Section \ref{Recon}. We conclude this paper in Section \ref{Con}.\

\section{Preliminaries}
\label{Preli}
\subsection{Fractional Fourier transform}
\label{Fractional}
\begin{defn} The fractional Fourier transform (FRFT) of a signal $x(t)$ with an angle $\alpha $ is defined as \cite{L. B. Almeida1994}

\begin{eqnarray}
\begin{aligned}
X_{\alpha}(u)=\mathcal{F}_{\alpha}\left[x(t) \right](u) \triangleq\int_{-\infty}^{+\infty}x(t)K_{\alpha}(u,t)dt,
\label{1}
\end{aligned}
\end{eqnarray}
where $\mathcal{F}_{\alpha}$ is the FRFT operator, $u$ stands for fractional frequency,  $K_{\alpha}(u,t)$ denotes the kernel function of the FRFT 
\begin{eqnarray}
\begin{aligned}
	K_{\alpha}(u,t)=
	\begin{cases}
		A_{\alpha}e^{j(\frac{\cot\alpha}{2} t^2-\csc\alpha ut+\frac{\cot\alpha}{2}u^2)},  &\alpha \neq k\pi\\
		\delta(t-u),       & \alpha=2k\pi\\
		\delta(t+u),      &\alpha=(2k-1)\pi
	\end{cases}
\label{2}		
\end{aligned}
\end{eqnarray}
where $A_\alpha\triangleq\sqrt{\frac{1-j\cot\alpha}{2\pi}}$, the rotation angle of FRFT is expressed as $\alpha=\dfrac{ p\pi}{2}$ and $p$ is the order of FRFT. The domain $0<\alpha<\dfrac{\pi}{2}$ are called fractional Fourier domains in \cite{Fractional1995}, and this definition is also adopted in this paper. 
\end{defn}

The FRFT can be understood as the rotation of the time-frequency plane. The essence of the FRFT of a signal is to decompose the signal with the chirp signal $K_{\alpha}(u,t)$ as the basis function. According to the FRFT of the signal $x(t)$, it can be determined whether it is bandlimited in the fractional domain.\

The FRFT has linear transform additivity, namely
\begin{eqnarray}	
\mathcal{F}_{\alpha+\beta}\left[x(t) \right](u) =\mathcal{F}_{\alpha}\left[x(t) \right](u)\cdot\mathcal{F}_{\beta}\left[x(t) \right](u)=	X_{\alpha}(u)\cdot	X_{\beta}(u).
\label{3}
\end{eqnarray}
It can be seen that the inverse transform of the FRFT relative to the $\alpha$ angle is the FRFT with the parameter $-\alpha$ angle, we have 
\begin{eqnarray}	
x(t)=\mathcal{F}_{-\alpha}\left\lbrace X_{\alpha}(u)\right\rbrace =\int_{-\infty}^{+\infty}X_{\alpha}(u)K_{-\alpha}(u,t)du,
\label{4}
\end{eqnarray}
when $\alpha=-\dfrac{\pi}{2}$, the FRFT degenerates to the traditional inverse FT; when $\alpha=\dfrac{\pi}{2}$, the FRFT degenerates to traditional FT, $X_{\frac{\pi}{2}}(u)=\int_{-\infty}^{+\infty}x(t)e^{-j2\pi ut}dt$; when $\alpha=0$, the FRFT degenerates to an identity transformation, $X_{0}(u)=x(t)$; when $\alpha=\pi$, the FRFT degenerates to the inversion of the signal with respect to the time axis, $X_{\pi}(u)=x(-t)$.\ 

\begin{defn} A signal $x(t)$ is called $\Omega_{\alpha}$ bandlimited signal in the FRFT, which means 
\begin{eqnarray}	
X_{\alpha}(u)=0, \qquad \mid u\mid>\Omega_{\alpha},
\label{5}
\end{eqnarray}
where $\Omega_{\alpha}$ is called the bandwidth of signal $x(t)$ in the FRFD. It has been shown that if the signal is bandlimited in the $\alpha$th FRFD, it can't be bandlimited in the FRFT with another angle $\beta$, where $\beta\neq\pm\alpha+n\pi$ for any integer $n$ \cite{X. G. Xia1996}.\
\end{defn}
	
\subsection{Unlimited sampling theorem in the Fourier domain}
\label{domain}
\begin{defn} The central modulo operation is defined by the mapping \cite{On2017}
\begin{eqnarray}
\begin{aligned}
	\mathscr{M}_{\lambda}: g\longmapsto2\lambda\left( \left[ \left[ \dfrac{g}{2\lambda}+\dfrac{1}{2}\right]\right]  -\dfrac{1}{2}\right),\qquad\left[ \left[ g\right] \right] \stackrel{def}{=}g-\lfloor g\rfloor, 
    \label{6}
\end{aligned}
\end{eqnarray}
where $\left[ \left[ g\right] \right]$ defines the fractional part of $g$ and $ \lambda>0$ is the ADC threshold. Note that Eq. (\ref{6})  is a nonlinear modulus mapping, which converts a smooth function into a discontinuous function. It is equivalent to a centered modulo operation since $\mathscr{M}_{\lambda}(g)\equiv g\:$ mod $\:2\lambda$. By implementing the mapping Eq. (\ref{6}), it is clear that out of range amplitudes are folded back into the dynamic range $[-\lambda,\lambda]$.
\end{defn}

Let's review some important conclusions in \cite{A.Bhandari2020,A.Bhandari2021}.
\begin{lem}\label{lem1} (Modular decomposition property) \cite{A.Bhandari2020} Let $g\in\mathcal{B}_{\Omega}$ where $\mathcal{B}_{\Omega}$ denotes the space of $\sigma$-bandlimited functions, and $\mathscr{M}_{\lambda}(\cdot)$ be defined in  Eq. (\ref{6}) where  $\lambda$ is a fixed, positive constant. Then, the bandlimited function $g(t)$ admits a decomposition
\begin{eqnarray}
\begin{aligned}
	g(t)=z(t)+\varepsilon_{g}(t),
	\label{7}
\end{aligned}
\end{eqnarray}
where $z(t)=\mathscr{M}_{\lambda}(g(t))$ and $\varepsilon_{g}(t)$ is a simple function
\begin{eqnarray}
\begin{aligned}
	\varepsilon_{g}(t)=2\lambda\sum\nolimits_{m\in\mathbb{Z}}e\left[ m\right]\oldstylenums{1}_{\mathcal{D}_{m}}(t), \quad e\left[m \right]\in\mathbb{Z},
	\label{8}  
\end{aligned}
\end{eqnarray}
where $\cup_{m\in\mathbb{Z}}\mathcal{D}_{m}=\mathbb{R}$ is a partition of the real line into intervals $\mathcal{D}_{m}$.
\end{lem}
The process of solving discontinuities is very critical. Lemma \ref{lem1} just proves this problem. Each bandlimited function, whether continuous or discrete, can be decomposed into the sum of the modular function and the stepwise residual of the simple function. Observe that the output function $z(t)$ is the difference between $g(t)$ and a piecewise constant function $\varepsilon_{g}(t)$.\                              
\begin{thm}(Unlimited sampling theorem in Fourier domain) \cite{A.Bhandari2021} Let $g\in\mathcal{B}_{\Omega}$ be a $\tau$-periodic function. Suppose that we are given $Q$ modulo samples of where $y[k]=\widetilde{\mathscr{M}_{\lambda}}(g(kT))$ folded at most $M$ times. Then a sufficient condition for recovery of $g(t)$ from $y[k]$ (up to a constant) is that, $T\leq\dfrac{\tau}{Q}$ and $Q\geq2\left( \dfrac{\Omega\sigma}{2\pi}+M+1\right)$.
\end{thm}
Different from the traditional FT which uses a complex exponential signal as the basis function, the FRFT uses a chirp signal as the basis function. This connotation determines that a non-bandlimited signal in the FD may be bandlimited in the FRFD. Therefore, the unlimited sampling study of non-bandlimited signals in the FD can be transformed into the theoretical study of bandlimited signals in the FRFD. The next step is to study the unlimited sampling theory of bandlimited signals in the FRFD.

\section{Mathematical Model for Unlimited Sampling}
\label{Mathematical}

In this Section, we will study the unlimited sampling theory of bandlimited signals in the FRFD. Here we make the following symbolic regulations: the sets of real, integer, and complex-valued numbers are denoted by $\mathbb{R}$, $\mathbb{Z}$, and $\mathbb{C}$, respectively. \

\subsection{Mathematical signal model}
\label{mapping}
Based on the periodic signal model in the FD, this paper proposes a periodic signal model in the FRFD, so mathematically $x(t)$ can be represented as follows. 

Let $\Omega_{\alpha}$ bandlimited square-integrable function $x(t)\in L^{2}$ satisfy $x(t)=x(t+\sigma)$, $\forall \sigma\in\mathbb{R}$. Then $x(t)$ has a fractional Fourier series expansion  
	\begin{eqnarray}
		\begin{aligned}
			x(t)&=\sum\limits_{|w|\leq R}\widehat{X}_{\alpha}(w)\Phi_{-\alpha}(w,t),
			\label{9}
		\end{aligned}
	\end{eqnarray}
	where $\widehat{X}_{\alpha}(w)$ is FRFS coefficient and 
	\begin{eqnarray}
		\begin{aligned}
			\Phi_{\alpha}(w,t)=\sqrt{\frac{\sin\alpha-j\cos\alpha}{\sigma}}e^{j(\frac{\cot\alpha}{2} t^{2}-\csc\alpha w u_{0}t+\frac{\cot\alpha}{2}w^{2}u_{0}^{2})},\label{10} 
		\end{aligned}
	\end{eqnarray}
	here, $R=\left\lceil\dfrac{\Omega_{\alpha}}{u_{0}}\right\rceil$, $u_{0}=\dfrac{2\pi\sin\alpha}{\sigma}$.

The discrete time representation $x(nT_{s}), n\in\mathbb{Z}$ of the signal $x(t)$ can be obtained by uniformly sampling at intervals of $T_{s}$. The discrete-time FRFT of the $\alpha$ angle of the discrete-time signal $x(t)$ is defined as follows
\begin{eqnarray}
	X_{\alpha,s}=\mathcal{F}_{\alpha}\left[x(nT_{s}) \right](u) \triangleq\sum_{n=-\infty}^{+\infty}x(nT_{s})K_{\alpha}(u,nT_{s}),
	\label{11}
\end{eqnarray}
where $K_{\alpha}$ is given by Eq. (\ref{2}), and $T_{s}$ is the sampling period.\

Let sampling function $x(t)\in L^{2}$ obtains $Q$ modular samples at the sampling rate $T$ in the interval $ t\in\left[-\frac{\sigma}{2},\frac{\sigma}{2}\right]$, then FRFS coefficients of $x(t)$ in the FRFD have a form   
	\begin{eqnarray}
		\widehat{X}_{\alpha}(w)= \left \{
		\begin{array}{ll}
			\int_{-\frac{\sigma}{2}}^{\frac{\sigma}{2}}x(t)\Phi_{\alpha}( w,t)dt,                  &
			w\in\mathbb{E}_{R,Q},
			\quad \\                  0,         & w\in\mathbb{I}_{Q}\backslash\mathbb{E}_{R,Q}, |w|>\Omega_{\alpha},
		\end{array}
		\right.
		\label{12}
	\end{eqnarray}
	where the set $\mathbb{I}_{Q}=\left\lbrace 0, 1, \cdots, Q-1 \right\rbrace$ denote the set of $Q$ contiguous integers, and $\mathbb{E}_{R,Q}$ is given by
	\begin{eqnarray}
		\mathbb{E}_{R,Q}=\left[0,R \right]\cup\left[Q-R,Q-1 \right],|\mathbb{E}_{R,Q}\mid=2R+1. 
		\label{13} 
	\end{eqnarray}

\begin{remark}The well-known Fourier series (FS) is just a special case of FRFS for $\alpha=\dfrac{\pi}{2}$, please see \cite{A.Bhandari2021}. In order to solve for $\widehat{X}_{\alpha}(w)$ in (\ref{12}), we must require $Q\geq2R+1$.
	Because of $QT=\sigma$, so $T\leq \dfrac{\sigma}{Q}\leq\dfrac{\sigma}{2R+1}.$\
\end{remark}

The hypothesis of periodic functions in our paper only provides a practical method for recovering signals from the folding measurements below. However, when the signal is aperiodic, the theoretical reconstruction guarantees that the aperiodic signal can also be expanded by discrete-time FRFT, but additional requirements are required for sampling samples, and this article will not expand in detail.

\subsection{Nonlinear modulus mapping}
\label{mapping}

This paper uses the definition and properties of generalized modular non-linear mapping in Eq. (\ref{6}), this phenomenon is equivalent to modulo arithmetic on the input function. And according to Lemma \ref{lem1}, gives the following form         
\begin{eqnarray}
		\begin{aligned}
			v_{x}(t)=x(t)-\mathscr{M}_{\lambda}\left( x(t)\right)=:\sum\nolimits_{m\in\mathbb{Z}}c\left[m\right]\mathbf{1}_{[t_{m},t_{m+1}]}\left(t \right),
			\label{14}  	 
		\end{aligned}
\end{eqnarray}
where $c\left[ m\right] \in\mathbb{R}$, $\mathbf{1}_{[t_{a},t_{b}]}$is the indicator function on $[t_{a},t_{b}]$, and $t_{m}\in\left[ -\frac{\sigma}{2},\frac{\sigma}{2}\right]$ denotes the folding instants with $t_{a}< t_{b}$. Obviously, the output function $\mathscr{M}_{\lambda}\left( x(t)\right)$ is the difference between $g(t)$ and a residual function $v_{x}(t)$. The \cite{A.Bhandari2020} requires that the correlation coefficient of the residual function $v_{x}(t)$ is an integer multiple of $2\lambda$, while \cite{A.Bhandari2021} doesn't need to make assumptions about its correlation coefficient. \

Without loss of the generality, we make the following symbolic regulations:
\begin{enumerate}[1)]
	\item Let $f\left[k \right]\overset{\mathrm{def}}{=}x\left( kT\right)$, $h\left[k \right] \overset{\mathrm{def}}{=}\mathscr{M}_{\lambda}\left( x\left( kT\right)\right)$, $v\left[k \right]\overset{\mathrm{def}}{=}v_{x}\left( kT\right)$, then
\begin{eqnarray}
	f\left[k \right] =h\left[k \right]+	v\left[k \right].
	\label{15}
\end{eqnarray}
  \item Let $\Delta^{N}f=\Delta^{N-1}\left( \Delta f\right)$ denote the $N$th difference operator with $\Delta f=f(k+1)-f(k)$, 
	$\overline{f}[k]\overset{\mathrm{def}}{=}\Delta f[k]$,
	$\overline{h}[k]\overset{\mathrm{def}}{=}\Delta h[k]$,
and $\overline{v}[k]\overset{\mathrm{def}}{=}\Delta v_{x}(kT)$, then 
\begin{eqnarray} 
	\begin{aligned}
		\overline{f}[k]&=\overline{h}[k]+\overline{v}[k]=\overline{h}[k]+\sum\nolimits_{m\in M}c\left[m\right]\delta(kT-t_{m}),\quad k\in\mathbb{I}_{Q}, 
		\label{16}
	\end{aligned}
\end{eqnarray}
where $\delta$ denotes the Dirac distribution, $c[m]$ are unknown weights, $t_{m}$ are unknown fold instant, and the size of the set $M$ depends on the dynamic range of the signal relative to the threshold $\lambda$. \
\end{enumerate}

Similar to the phase unwrapping theory, we can obtain the following fact from Itoh’s condition \cite{Itoh1982}. when the max-norm of the first-order finite difference of the samples is bounded by $2\lambda$ or $|f[k+1]-f[k]|\leq 2\lambda$, the first-order finite difference operator on the modular sequence can be reversed operation to restore. \

Eq. (\ref{16}) is written as the FRFD
\begin{eqnarray}
\overline{H}_{\alpha}[n]= \left \{
\begin{array}{ll}
	\overline{F}_{\alpha}[n]-\overline{V}_{\alpha}[n],                  & n\in\mathbb{E}_{R,Q-1}\\
	-\;\overline{V}_{\alpha}[n],   & n\in\mathbb{I}_{Q-1}\backslash\mathbb{E}_{R,Q-1}\\
\end{array}
\right.
  \label{17}
\end{eqnarray}
where $\overline{F}_{\alpha}$, $\overline{H}_{\alpha}$, and $\overline{V}_{\alpha}$ are the FRFT of $\overline{f}$, $\overline{h}$, and $\overline{v}$ respectively. At the same time, the discrete FRFT form of $\overline{h}[k]$ is given
\begin{eqnarray}
\begin{aligned}
	\overline{H}_{\alpha}[n]&=\sum\nolimits_{k\in\mathbb{I}_{Q-1}}\overline{h}[k]K_{\alpha}(n,k)\\
	&=\sum\nolimits_{k\in\mathbb{I}_{Q-1}}A_{\alpha}\overline{h}[k]\,e^{j\left( \frac{\cot\alpha}{2}k^{2}-\csc\alpha \overline{u}_{0}kn+\frac{\cot\alpha}{2} \overline{u}_{0}^{2}n^{2}\right) },
	\label{18}
\end{aligned}
\end{eqnarray}
where $A_\alpha\triangleq\sqrt{\frac{1-j\cot\alpha}{2\pi}}$ and $\overline{u}_{0}=\dfrac{2\pi\sin\alpha}{Q-1}$. When $\alpha=\dfrac{\pi}{2}$, this transform is the discrete FT, see \cite{A.Bhandari2021} for details.\

If we want to recover $f[k]$, we must solve $v[k]$, then Eq. (\ref{15}) is transformed into solving Eq. (\ref{16}), and the key to solving Eq. (\ref{17}) is to find the value of the unknown folding instant $\left\lbrace c[m],t_{m}\right\rbrace_{m\in\mathbb{Z}}$. Using Eq. (\ref{1}), we can obtain 
\begin{eqnarray}
\begin{aligned}
	\overline{V}_{\alpha}[n]&=\sum\nolimits_{k\in\mathbb{I}_{Q-1}}\sum\nolimits_{m\in M}c[m]\delta\left(kT-t_{m} \right) \Phi_{\alpha}(n,kT)\\
	&=\sum\nolimits_{m\in M}c[m]A_{\alpha}e^{j\left(\frac{\cot\alpha }{2T^{2}}t_{m}^{2}-\frac{\csc\alpha \overline{u}_{0}n}{T}t_{m}+\frac{\cot\alpha}{2} \overline{u}_{0}^{2}n^{2} \right) },
	\label{19}
\end{aligned}
\end{eqnarray}
where $A_\alpha\triangleq\sqrt{\frac{1-j\cot\alpha}{2\pi}}$, $M=|\mathcal{M}|$. When $\alpha=\dfrac{\pi}{2}$, this transform is the discrete FT \cite{A.Bhandari2021}. The estimation of the unknown parameters in Eq. (\ref{19}) is called the spectral estimation problem \cite{P. Stoica2000,Spectral2005}.\

\section{Unlimited sampling theorem in fractional Fourier domain}
\label{Recon}
\subsection{Computing the folding instants}
\label{theor}
If we want to recover $v[k]$, we must find the value of the unknown folding instant $\left\lbrace c[m],t_{m}\right\rbrace_{m\in\mathbb{Z}}$. Eq. (\ref{19}) is the spectral estimation problem. The commonly used spectrum estimation methods are annihilation filter(AF) \cite{P. Stoica2000, M Vetterli2002}, ESPRIT \cite{ESPRIT1989}, MUSIC \cite{TSTMUSIC2001}, etc. Among them, AF is the most commonly used method in many theoretical analyses and practical applications. In principle, the signal reconstruction process is to use the obtained set of moments or fractional Fourier coefficients of the input signal to solve a spectrum problem to achieve an accurate estimation of the unknown parameter $\left\lbrace c[m], t_{m}\right\rbrace_{m\in\mathbb{Z}}$. For convenience, Eq. (\ref{19}) is written as follows
\begin{eqnarray}
\begin{aligned}
	\overline{V}_{\alpha}[n]=A_{\alpha}\underbrace{e^{j\frac{\cot\alpha}{2} \overline{u}_{0}^{2}n^{2}}}_{\kappa(n)}\left(\underbrace{\sum\nolimits_{m\in M} \underbrace{c[m]e^{j\frac{\cot\alpha}{2T^{2}} t_{m}^{2}}}_{\chi_{m}}\cdot \underbrace{e^{-j\csc\alpha \frac{\overline{u}_{0}}{T}nt_{m}}}_{\varsigma_{m}^{n}}}_{\Im(n)}\right),
	\label{20}  
\end{aligned}
\end{eqnarray}
where $\Im(n)=\sum\nolimits_{m\in M}\chi_{m}\varsigma_{m}^{n}$. Since the formula is very complicated, we rewrite the formula  $\overline{V}_{\alpha}[n]=A_{\alpha}\kappa(n)\Im(n)$. Because the part of $\kappa(n)$ does not contain unknowns parameters, we separately perform an annihilation filter to get $\left\lbrace c[m], t_{m}\right\rbrace$, and finally bring in Eq. (\ref{20}), and get $\overline{v}$  through inverse FRFT.\

First, let's analyze $\Im(n)$ in detail below, 
\begin{eqnarray}
\begin{aligned}
    \Im(n)=\sum\nolimits_{m\in M}\chi_{m}\varsigma_{m}^{n}.
    \label{21}
\end{aligned}
\end{eqnarray}
Eq. (\ref{21}) is a classic spectrum estimation problem, which can be handled by an annihilation filter. It is known from the \cite{P. Stoica2000} that we can accurately estimate the unknown parameters $\chi_{m}$ and $\varsigma_{m}$ from $2K$ continuous non-zero measured values $\Im(n)$. The following is divided into two parts to solving separately.\

(1) Construct the filter $\left\lbrace \Gamma[\vartheta]\right\rbrace_{\vartheta=0,1,\cdots,M} $ so that its zero point is the parameter $\varsigma_{m}=\left\lbrace e^{-j\frac{\csc\alpha \overline{u}_{0}}{T}t_{\vartheta}}\right\rbrace_{\vartheta=0}^{\vartheta=M-1}$, then the $z$ transform of the filter can be expressed as
\begin{eqnarray}
\begin{aligned}
	\Gamma[z]=\prod_{m=0}^{M-1}(1-\varsigma_{m}z^{-1})=\sum_{\vartheta=0}^{M}\Gamma[\vartheta]z^{-\vartheta}.
	\label{22}
\end{aligned}
\end{eqnarray} 
It can be seen that the root of the polynomial is the parameter $\varsigma_{m}$. Therefore, this paper convolutes it directly, so it has
\begin{eqnarray}
\begin{aligned}
	\left(\Gamma\ast\Im\right)[n]&=\sum_{\vartheta=0}^{M}\Gamma[\vartheta]\Im[n-\vartheta]\\
	&=\sum_{\vartheta=0}^{M}\sum_{m=0}^{M-1}c[m]e^{j\frac{\cot\alpha}{2T^{2}}t_{m}^{2}}\cdot\Gamma[\vartheta]\cdot e^{-j\frac{\csc\alpha (n-\vartheta)\overline{u}_{0}}{T}t_{m}}\\
	&=\sum_{m=0}^{M-1}c[m]e^{j\frac{\cot\alpha}{2T^{2}}t_{m}^{2}}\underbrace{\sum_{\vartheta=0}^{M}\Gamma[\vartheta]e^{j\frac{\csc\alpha\vartheta \overline{u}_{0}}{T}t_{m}}}_{\Gamma[\varsigma_{m}]}e^{-j\frac{\csc\alpha n\overline{u}_{0}}{T}t_{m}}=0.
	\label{23}
\end{aligned}
\end{eqnarray} 
We write Eq. (\ref{23}) in the form of matrix vector to obtain
\begin{eqnarray}
\begin{aligned}	
	\begin{bmatrix}
		\Im[M-1]  &  \Im[M-2]    &  \cdots   &  \Im[0]\\
		\Im[M]  & \Im[M-1]     &\cdots     &\Im[1]\\
		\vdots  & \vdots     &\vdots     &\vdots\\
		\Im[N-1]  &  \Im[N-2]    &\cdots     &\Im[N-M] 
	\end{bmatrix}
	\begin{bmatrix}
		\Gamma[1]\\
		\Gamma[2]\\
		\vdots\\
		\Gamma[M]
	\end{bmatrix}
	=-	
	\begin{bmatrix}
		\Im[M]\\
		\Im[M+1]\\
		\vdots\\
		\Im[N]
	\end{bmatrix},
  \label{24}
\end{aligned}
\end{eqnarray} 
where $\Im=\big[\:\Im[0], \Im[1], \cdots, \Im[M]\:\big] ^{T}$, $\Im[M]=1$. The unique solution a can be obtained $\Gamma[\vartheta],  \vartheta=1, 2, \cdots, M $, and finally, the instantaneous folding time $\left\lbrace t_{m}\right\rbrace_{m\in\mathbb{Z}}$ can be obtained.\

(2) In order to estimate the amplitude parameter $\chi_{m}$, extract $M$ continuous values from the known coefficient $\Im[n]$, that is to say, $m=0, 1, \cdots, M$, and write the 
$\Im(n)=\sum_{m=0}^{M-1}\chi_{m}\varsigma_{m}^{n}$
in the form of a matrix-vector
\begin{eqnarray}
\begin{aligned}
	U\chi=\Im,
	\label{25}
\end{aligned}
\end{eqnarray}
\begin{eqnarray}
\begin{aligned}\label{FRFT}
	\begin{bmatrix}
		1              &     1    &\cdots           & 1\\
		\varsigma_0           &    \varsigma_1    &\cdots         &\varsigma_{M-1}\\
		\vdots       & \vdots      &  \vdots      &\vdots\\
		\varsigma_0^{M-1}   &\varsigma_1^{M-1}      &\cdots         &\varsigma_{M-1}^{M-1}
	\end{bmatrix}
	&\begin{bmatrix}
		e^{j\frac{\cot\alpha}{2T^{2}} t_{0}^{2}}       &    \     &\           & \  \\
		\ &  e^{j\frac{\cot\alpha}{2T^{2}} t_{1}^{2}}      &      &\         &\   \\
		\       & \       &  \ddots     & \    \\
		\    &\      &\         & e^{j\frac{\cot\alpha}{2T^{2}} t_{M-1}^{2}}
	\end{bmatrix}\\
	&\cdot\begin{bmatrix}
		c[0]\\
		c[1]\\
		\vdots\\
		c[m-1]	
	\end{bmatrix}	
	=
	\begin{bmatrix}
		\Im[0]\\
		\Im[1]\\
		\vdots\\
		\Im[M-1]
	\end{bmatrix}.
  \label{26}
\end{aligned}
\end{eqnarray}
where $U$ is a vandermonde matrix, it is a matrix whose columns are geometric series. For any integer $a, b=0, 1, \cdots, M-1$ ($a\neq b$) satisfy $U_{a}\neq U_{b}$, and $U$ is non-singular, at this time, Eq. (\ref{25}) has a unique solution. It needs to be emphasized here that we generally use the least squares method to obtain an estimate of the amplitude information.\

Through the above two steps, the instantaneous folding time $\left\lbrace c[m],t_{m}\right\rbrace$ can be obtained. we bring $\left\lbrace c[m],t_{m}\right\rbrace$ into  Eq. (\ref{20}), and get $\overline{v}$  through inverse fractional Fourier transform.

\subsection{Unlimited sampling theorem in the fractional Fourier domain}
\label{fract}
Through the above research content, we have found the value of the unknown folding instant $\left\lbrace c[m],t_{m}\right\rbrace_{m\in\mathbb{Z}}$. If $v[k]$ is known, we can infer $v[k]$ from $h[k]$ and recover $f[k]$ from $h[k]$. The unlimited sampling theorem in the FRFT domain is given below.
\begin{thm}\label{lem2} (Unlimited sampling theorem in FRFT domain) Let $\Omega_{\alpha}$ bandlimited square-integrable function $x(t)\in L^{2}$ satisfy $x(t)=x(t+\sigma)$, $\forall \sigma\in\mathbb{R}$, and $h\left[k \right] =\mathscr{M}_{\lambda}\left( x\left( kT\right)\right)$ folded at most $M$ times. Then a sufficient condition for recovery of $x(t)$ from $h[k]$, is that $T\leq\dfrac{\sigma}{Q}$ and $Q\geq2\left( \dfrac{\Omega_{\alpha}\sigma}{2\pi\sin\alpha}+M+1\right)$, where $M$ is known.
\end{thm}
Proof: From the unlimited sampling part of the fractional domain in Eq. (\ref{17}) and  Eq. (\ref{20}), we can get the
$$
\begin{bmatrix}
\Im[M-1]  &  \Im[M-2]    &  \cdots   &  \Im[0]\\
\Im[M]  & \Im[M-1]     &\cdots     &\Im[1]\\
\vdots  & \vdots     &\vdots     &\vdots\\
\Im[N-1]  &  \Im[N-2]    &\cdots     &\Im[N-M] 
\end{bmatrix}
\begin{bmatrix}
\Gamma[1]\\
\Gamma[2]\\
\vdots\\
\Gamma[M]
\end{bmatrix}
=-	
\begin{bmatrix}
\Im[M]\\
\Im[M+1]\\
\vdots\\
\Im[N]
\end{bmatrix},
$$ 
that is, $\Gamma[\vartheta]$. From $$\left(\Gamma\ast\Im\right)[n]=0.$$ We can get the root of $\varsigma_{m}$, bring Eq. (\ref{19}) into Eq. (\ref{17}), use the least square method to estimate $c[m]$, and get $v_{k}$ in Eq. (\ref{15}). we develop a method that allows for inferring $v[k]$ from $h[k]$. Finally, we get the sampling density criterion, and the following conclusions can be drawn
\begin{eqnarray}
	\begin{aligned}
		|\mathbb{I}_{Q-1}\backslash\mathbb{E}_{R,Q-1}|=K-2R-2\geq2M,
		\label{27}
	\end{aligned}
\end{eqnarray}
where $M$ is known. Due $QT=\sigma$, $R=\left\lceil\dfrac{\Omega_{\alpha}}{u_{0}}\right\rceil$, $u_{0}=\dfrac{2\pi\sin\alpha}{\sigma}$, we have
\begin{eqnarray}
	\begin{aligned}
		T=T_{FRFT}\leq\dfrac{\sigma}{2(R+M+1)}=\dfrac{\sigma}{2(\lceil\Omega_{\alpha}\sigma/2\pi\sin\alpha\rceil+M+1)}.
		\label{28}
	\end{aligned}
\end{eqnarray}
 Proof completed. $\hfill\qedsymbol$\
\begin{remark}
After performing $M$ folds, Eq. (\ref{28}) can guarantee the restoration and reconstruction of folding moment $\left\lbrace c[m],t_{m}\right\rbrace_{m=0}^{M-1}$.  Theorem \ref{lem2} turns out that unlimited sampling theorem has nothing to do with the modulus threshold and can handle arbitrary folding time. When $\alpha=\dfrac{\pi}{2}$, see \cite{A.Bhandari2021}.\
\end{remark}
The unlimited sampling theorem proves that the non-bandlimited signal in the FD based on the FRFT can be recovered from analog sampling as long as it meets Eq. (\ref{28}) whose amplitude exceeds the ADC threshold by orders of magnitude. Special emphasis the signal is not affected by the ADC threshold.\

\section{Conclusion}
\label{Con}
In this article, we study the sampling theorem of bandlimited signals in the fractional Fourier domain based on the unlimited sampling framework of modulo measurement. Our main work is to perform modular operations in the fractional Fourier domain with the folding introduced by modular nonlinearity, and then to deal with the problem of fractional spectrum estimation. It turns out that unlimited sampling theorem has nothing to do with the modulus threshold and can handle arbitrary folding time.\

\section*{Declaration of competing interest}
The authors declare that they have no known competing financial interests or
personal relationships that could have appeared to influence the work reported in this paper.



\

\

\
%


\begin{thebibliography}{999}
\bibitem{Communication1949} C. E. Shannon, Communication In The Presence Of Noise, Proc. Ire. 37 (1) (Jan. 1949) 447-457.
\bibitem{Certain2002} H. Nyquist, Introduction to Certain topics in telegraph transmission theory, Proc. IEEE 90 (2) (Mar. 2002) 276-279.
\bibitem{Sampling1992} P. L. Butzer, R. L. Stens, Sampling theory for not necessarily band-limited functions: a historical overview, SIAM Review. 34 (1) (Mar. 1992) 40–53.
\bibitem{Unser2000} M. Unser, Sampling–50 years after Shannon, Proc. IEEE 88 (4) (Apr. 2000) 567–587.
\bibitem{Quantitative1999} T. Blu, M. Unser, Quantitative Fourier analysis of approximation techniques. I. Interpolators and projectors, IEEE Trans. Signal Process. 47 (10) (Oct. 1999) 2796–2806.
\bibitem{Higher2019} R. M. Jing, Q. Feng, B. Z. Li, Higher order derivative sampling associated with fractional Fourier	transform, Circuits Syst. Signal Process. 38 (11) (Apr. 2019) 1751–1774.
\bibitem{New2007} B. Z. Li, R. Tao, Y. Wang, New sampling formula related to linear canonical transform, Signal Process. 87 (5) (May. 2007) 983–990.  
\bibitem{Seven2014} P. L. Butzer, M. M. Dodson, P. J. S. G. Ferreira, J. R. Higgins, G. Schmeisser, R. L. Stens, Seven pivotal theorems of Fourier analysis, signal analysis, numerical analysis and number theory: their interconnections, Bull. Math. Sci. 4 (3) (Dec. 2014) 481–525.	\bibitem{Amplitude2009} U. K. Kwon, D. Kim, G. H. Im, Amplitude clipping and iterative reconstruction of MIMO-OFDM signals with optimum equalization, IEEE Trans. Wirel. Commun. 8 (1) (Jan. 2009) 268–277.
\bibitem{Iterative2014} Z. Koll\'{a}r, J. Gazda, P . Horv\'{a}th, L. Varga, D. Kocur, Iterative signal reconstruction of deliberately clipped SMT signals, Sci. China Inf. Sci. 56 (2) (Feb. 2014) 1–13.
\bibitem{Restoring1991} J. S. Abel, J. O. Smith, Restoring a clipped signal, in: International Conference on Acoustics, Speech, and Signal Processing (ICASSP), April 14-17, 1991, pp. 1745–1748.	
\bibitem{W. Kester2009} W. Kester, MT-025 tutorial ADC architectures VI: folding ADCs, Analog Devices Tech. Rep. 27 (2014) 623–656.
\bibitem{D. Park2007} D. Park, J. Rhee, Y. Joo, A wide dynamic-range CMOS image sensor using self-reset technique, IEEE Electron Device Lett. 28 (10) (Oct. 2007) 890–892.
\bibitem{Y. Joo2010} D. Park, Y. Joo, S. Koppa, A simple and robust self-reset CMOS image sensor, in: International Midwest Symposium on Circuits and Systems (MWSCAS), August 1-4, 2010, pp. 680–683.
\bibitem{On2017} A. Bhandari, F. Krahmer, R. Raskar, On unlimited sampling, in: International Conference on Sampling Theory and Applications (SampTA), July 3-7, 2017, pp. 1-26.
\bibitem{A.Bhandari2020} A. Bhandari, F. Krahmer, R. Raskar, On unlimited sampling and reconstruction, IEEE Trans. Signal Process. 69 (Dec. 2020) 3827-3839.
\bibitem{Shah2018} V. Shah, C. Hegde, Signal reconstruction from modulo observations, in: Global Conference on Signal and Information Processing (GlobalSIP), November 11-14, 2019, pp. 1-5.
\bibitem{A.Bhandari2018} A. Bhandari, F. Krahmer, R. Raskar, Unlimited sampling of sparse sinusoidal mixtures, in: International Symposium on Information Theory (ISIT), June 17-22, 2018, pp. 336-340. 
\bibitem{Sparse2019} L. Rencker, F. Bach, W. Wang, M. D. Plumbley, Sparse recovery and dictionary learning from nonlinear compressive measurements, IEEE Trans. Signal Process. 67 (21) (Nov, 2019) 5659-5670.
\bibitem{Recovering2019} F. Ji, Pratibha, W. P. Tay, Recovering graph signals from folded samples. (Jan. 2020) 1-30. arXiv:1903.03741v2.	\bibitem{Above2019} E. Romanov, O. Ordentlich, Above the Nyquist rate, modulo folding does not hurt, IEEE Signal Process. Lett. 26 (8) (Aug. 2019) 1167–1171. 
\bibitem{Ordentlich2018} O. Ordentlich, G. Tabak, P. K. Hanumolu, A. C. Singer, G. W. Wornell, A modulo-based architecture for analog-to-digital conversion, IEEE J. Sel. Top. Signal Process. 12 (5) (Aug. 2018) 825-840.
\bibitem{Generalized2018} O. Musa, P. Jung, N. Goertz, Generalized approximate message passing for unlimited sampling of sparse signals, in: Global Conference on Signal and Information Processing (GlobalSIP), November 26-29, 2018, pp. 336-340. 
\bibitem{S. Rudresh2018} S. Rudresh, A. Adiga, B. A. Shenoy, C. S. Seelamantula, Wavelet-based reconstruction for unlimited sampling, in: International Conference on Acoustics, Speech and Signal Processing (ICASSP), April 15-20, 2018, pp. 4584-4588. 
\bibitem{A.Bhandari2021} A. Bhandari, F. Krahmer, T. Poskitt, Unlimited sampling from theory to practice: Fourier-Prony recovery and prototype ADC, IEEE Trans. Signal Process. 70 (Sep. 2021) 1131-1141.
\bibitem{X. G. Xia1996} X. G. Xia, On bandlimited signals with fractional Fourier transform, IEEE Signal Process. Lett. 3 (3) (Mar. 1996) 72–74. 
\bibitem{L. B. Almeida1994} L. B. Almeida, The fractional Fourier transform and time-frequency representation, IEEE Trans. Signal Process. 42 (11) (Nov. 1994) 3084–3091.
\bibitem{Fractional1995} H. M. Ozaktas, O. Aytür, Fractional Fourier domains, Signal Process. 46 (1) (Sep. 1995) 119–124.
\bibitem{Pei1999} S. C. Pei, M. H. Yeh, T. L. Luo, Fractional Fourier series expansion for finite signals and dual extension to discrete-time fractional Fourier transform, IEEE Trans. Signal Process. 47 (10) (Oct. 1999) 2883–2888.
\bibitem{Marziliano2009} A. Bhandari, P. Marziliano, Sampling and reconstruction of sparse signals in fractional Fourier domain, IEEE Signal Process. Lett. 17 (3) (Apr. 2010) 221-224.
\bibitem{Itoh1982} K. Itoh, Analysis of the phase unwrapping algorithm, Applied Optics. 21 (14) (Jul. 1982) 2470.
\bibitem{P. Stoica2000} P. Stoica, R. Moses, Introduction to spectral analysis. Englewood Cliffs, NJ: Prentice-Hall, 2000.
\bibitem{Spectral2005} P. Stoica, R. Moses, Spectral analysis of signals. London: Prentice Hall, 2005.
\bibitem{M Vetterli2002} M. Vetterli, P. Marziliano, T. Blu, Sampling signal with finite rate of innovation, IEEE Trans. Signal Process. 50(6) (Jun. 2002) 1417–1428.
\bibitem{ESPRIT1989} R. Roy, T. Kailath, Esprit-estimation of signal parameters via rotational invariance technique, IEEE Trans. Acoust. Speech. Signal Process. 37 (7) (Aug. 1989) 984–995.
\bibitem{TSTMUSIC2001} Y. Y. Wang, J. T. Chen, W. H. Fang, TST-MUSIC for joint DOA-delay estimation, IEEE Trans. Signal Process. 49 (4) (Apr. 2001) 721–729.

\end{thebibliography}
\end{document}